\documentclass[letter,12pt]{article}
\usepackage{todonotes}
\usepackage{newtxtext,newtxmath, amsfonts, bm,mathtools}
\mathtoolsset{showonlyrefs=true}
\usepackage[utf8]{inputenc}
\usepackage[boxed,lined,linesnumbered]{algorithm2e}
\usepackage{amsmath, textcomp, paralist, bm, natbib}
\usepackage{array}
\usepackage{wrapfig}
\usepackage{multirow}
\usepackage{tabularx}
\usepackage{booktabs}
\usepackage{float}
\DeclareMathAlphabet{\mathcal}{OMS}{cmsy}{m}{n}
\graphicspath{{figures/}}
\usepackage{arydshln}
\usepackage{xcolor}
\usepackage[hyphens,spaces,obeyspaces]{url}
\usepackage{indentfirst}

\usepackage[margin=3cm]{geometry}
\usepackage[margin=2cm]{caption}
\usepackage{titling}

\usepackage{tikz}
\usetikzlibrary{positioning}

\SetKwInOut{Parameter}{parameter}

\usepackage{adjustbox}
\newcolumntype{R}[2]{%
    >{\adjustbox{angle=#1,lap=\width-(#2)}\bgroup}%
    l%
    <{\egroup}%
}
\newcommand*\rot{\multicolumn{1}{R{90}{1em}}}

\usepackage{array, ragged2e}
\newcolumntype{P}[1]{>{\raggedright\arraybackslash}p{#1}}
\begin{document}
%
\title{Honeyboost: Boosting honeypot performance with data fusion and  anomaly detection }

\author{Sevvandi Kandanaarachchi \thanks{\scriptsize{Email:sevvandi.kandanaarachchi@rmit.edu.au,  Affiliation:School of Science (Mathematical Sciences), RMIT University, Melbourne VIC 3000, Australia.}}, Hideya Ochiai \thanks{\scriptsize{Email:ochiai@elab.ic.i.u-tokyo.ac.jp, Affiliation:Graduate School of  Information Science and Technology, The University of Tokyo, Tokyo, Japan}}, Asha Rao \thanks{\scriptsize{Email:asha.rao@rmit.edu.au, Affiliation:School of Science (Mathematical Sciences), RMIT University, Melbourne VIC 3000, Australia.}}}
\date{%
   \scriptsize{ \today} \\
}

\begin{titlingpage}
\maketitle





\begin{abstract}
With cyber incidents and data breaches becoming increasingly common, being able to predict a cyberattack has never been more crucial. The ability of Network Anomaly Detection Systems (NADS) to identify unusual behavior makes them useful in predicting such attacks. However, NADS often suffer from high false positive rates. In this paper, we introduce a novel framework called Honeyboost that enhances the performance of honeypot aided NADS. Using data from the LAN Security Monitoring Project, Honeyboost identifies most anomalous nodes before they access the honeypot aiding early detection and prediction. Furthermore, using extreme value theory, we achieve the highly desirable low false positive rates.

Honeyboost is an unsupervised method comprising two approaches:  horizontal and vertical. The horizontal approach constructs a time series from the communications of each node, with node-level features encapsulating their behavior over time. The vertical approach finds anomalies in each protocol space.  Using a window-based model, which is typically used in online scenarios, the horizontal and vertical approaches are combined to identify anomalies and gain useful insights. Experimental results indicate the efficacy of our framework in identifying suspicious activities of nodes.
\end{abstract}


\end{titlingpage}

\newgeometry{top=1.5cm,bottom=2cm,right=1.5cm,left=1.5cm}
\section{Introduction}\label{sec:intro}

Increasing cyber attacks and data breaches require new ways of predicting attacks, especially previously unseen ones. A very popular tool for prediction, is a honeypot \citep{Barak2020} that aims to lure attackers and learn their tactics. While honeypots have known advantages such as ease of installation and low  resource usage \citep{campbell2015survey}, there exist drawbacks, such as limited vision, discovery and fingerprinting, that increase the risk of takeover \citep{Mokube2007}. Thus, honeypots are often deployed in conjunction with Network Anomaly Detection Systems (NADS) \citep{BAYKARA2018103,7724682}. 

NADS are capable of detecting new attacks, a key requirement in cyber security. However, they suffer from high false positive rates. Furthermore, honeypot aided NADS are generally deployed in public domain networks. The use of honeypots in Local Area Networks (LANs) has not been widely explored. In this paper, we propose a novel framework, \textit{Honeyboost}, that enhances LAN honeypot performance with Network Anomaly Detection (NAD), while giving low false positive rates. Honeyboost identifies most anomalous nodes in the LAN, before they even access the honeypot, enabling better prediction of cyber attacks.





While many researchers have studied honeypots and anomaly detection, most place these detection tools in the public/global IP domain or in Internet gateway routers. This  results in efficient detection of global cyber-attack behaviors such as global-scans, Distributed Denial of Service (DDoS) attacks, or botnet constructions. However, such placing is inadequate for detecting local or LAN internal cyber-attack behaviors such as malware propagation, insider attacks, or data stealing through direct communications in the same network segment without passing through routers. This may happen when (1) a malware is pre-installed in a mobile device that connects to a WiFi, or (2) a malware is delivered to a computer through a phishing e-mail or Social Network Sites (SNS). 

This paper focuses on honeypot traffic in a local area network, where the behavior of the address resolution protocol (ARP) can be observed. ARP requests are normally broadcast in the local network segment in order to find the  media access control (MAC) address of the target IP address before forwarding an IP packet to the host. The sequences of ARP requests, along with other protocol data are used as input for anomaly detection. This allows suspicious behavior, such as multiple attempts to access many IP addresses in the local area network, to be detected as anomalous. This LAN based honeypot research is unique as it allows us to predict anomalous nodes, something that cannot happen in a global setting as the nodes wouldn't be known. 

We recognize that not all anomalies -- nodes that access the honeypot -- are malicious. In general, anomalies are mainly caused by  \begin{inparaenum} \item malware trying to intrude into other hosts or steal data from the network data storage/camera, \item security software equipped with features of basic vulnerability testing, and \item network operators  intentionally accessing computers in the LAN to check their status. \end{inparaenum} 





Honeyboost treats the output traffic of each node/host in a LAN as a time series, and computes node-level features describing the behavior of nodes over time.  To the best of our knowledge, this is the first study that computes node-level features by treating the output traffic of each node as a time series, in honeypot aided NADS. We use an unsupervised anomaly detection (AD) method called \textit{Lookout} that uses extreme value theory to detect anomalies and effective in minimizing false positives. Thus, the performance of the honeypot is enhanced by using a node-level formulation in conjunction with an AD method with low false positives.

\begin{figure}[!ht]
     \centering
    \includegraphics[scale=0.7]{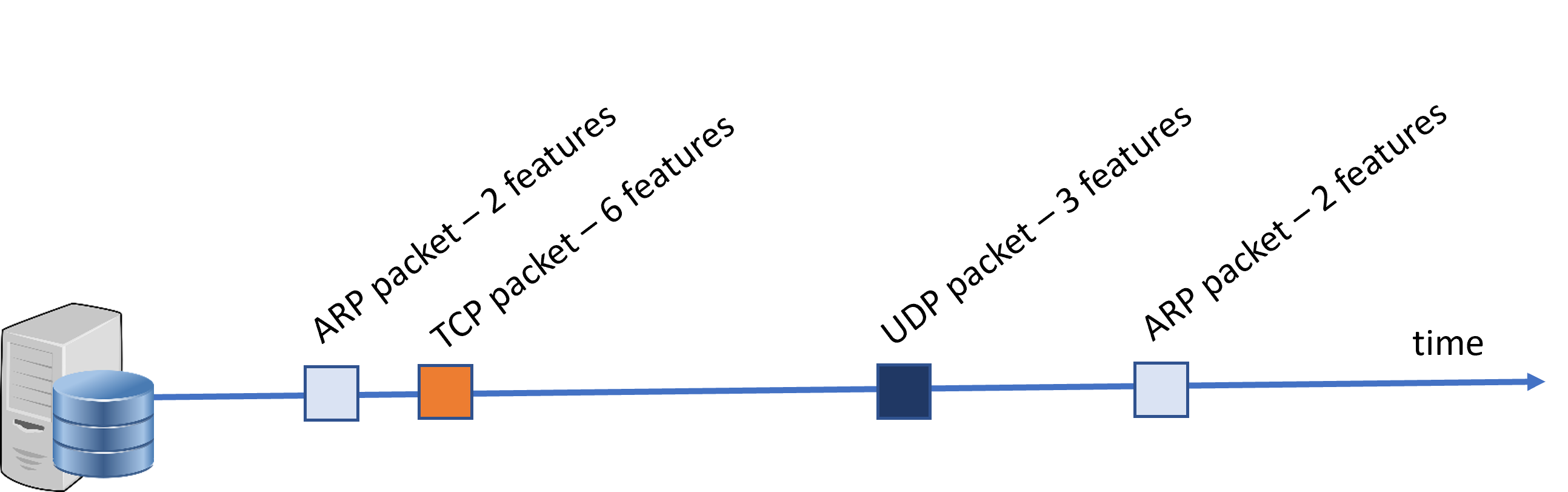}
     \caption{An example of  the output of a single node, which can be treated as a time series. As each protocol gives rise to a different number of features, this is a varying-dimensional time series.    }
     \label{fig:horizontalonenode}
 \end{figure}

Identifying anomalous nodes in a network can be formulated as a time series problem as nodes transmit network packets using multiple protocols at different time points (Figure~\ref{fig:horizontalonenode}). Thus, identifying anomalous time series would yield the nodes that behave quite differently from the rest.

\subsection{Challenges}
Formulating Network Anomaly Detection (NAD) as a time series problem, whether it is honeypot aided or not, presents several challenges: 
\begin{enumerate}
        \item \textit{Varying dimensional time series:} Univariate and multivariate time series are widely studied in the literature from many inter-disciplinary contexts. A \textit{univariate time series} is a sequence of real valued numbers indexed in time, denoted by $\{ u_1, u_2, \ldots, u_t, \ldots \}$, where $u_t \in \mathbb{R}$ for all $t$.  A \textit{multivariate time series} is a sequence of vectors indexed in time denoted by $\{ \bm{u}_1, \bm{u}_2, \ldots, \bm{u}_t, \ldots \}$, where $\bm{u}_t \in \mathbb{R}^p$ for all $t$. However, here we have a different situation. As the same node emits packets using different protocols, we have a time series $\{ \bm{u}_{t_1}, \bm{v}_{t_2}, \bm{w}_{t_3}, \ldots, \bm{u}_{t_i},\bm{v}_{t_j},\bm{w}_{t_k}, \ldots \}$, where $\bm{u}_{t_i} \in \mathbb{R}^p$, $\bm{v}_{t_j} \in \mathbb{R}^q$ and $\bm{w}_{t_k}\in \mathbb{R}^m$ for different values of $p, q$, $m$ and different times $t_i$, $t_j$ and $t_k$. We call such a time series \textit{varying-dimensional} (VD). There has not been much research conducted on VD time series. A literature search has not indicated any study of anomaly detection using VD time series. 
    \item \textit{Irregularity:} Unevenly spaced time series are referred to as irregular time series. The traditional time series theory is applicable to regular, evenly spaced time series, that is, a series $\{ u_1, u_2, \ldots, u_t, \ldots \}$, where the time between successive entries $u_i$ and $u_{i+1}$ is constant. Here we have an irregular varying-dimensional time series as the nodes communicate with each other at different time points. As such, when we consider the VD time series $\{ \bm{u}_{t_1}, \bm{v}_{t_2}, \bm{w}_{t_3}, \ldots, \bm{u}_{t_n}, \ldots \}$, in general $t_{n+1} - t_n \neq t_n - t_{n-1}$.
    \item  \textit{Sparse literature:} The literature on NADS mostly considers a packet based feature approach. We see this in most publicly available datasets. For example  popular datasets such as KDD Cup 99  and Kyoto dataset contain packet level observations without source details.  While we appreciate that source addresses may compromise privacy, without even a dummy identifier it is not possible to construct a time series at the node level. As such, we observe that while these datasets have been instrumental for the research growth in NADS, they have also steered the progress towards a packet based feature approach.  
\end{enumerate}
While a node-based VD time series approach to NAD has challenges, it takes a more holistic view on NAD.  As such, we expect to gain deeper insights about network anomalies by using this modeling paradigm.

\subsection{Our contribution}
\begin{figure*}[!ht]
     \centering
    \includegraphics[width=0.9\textwidth]{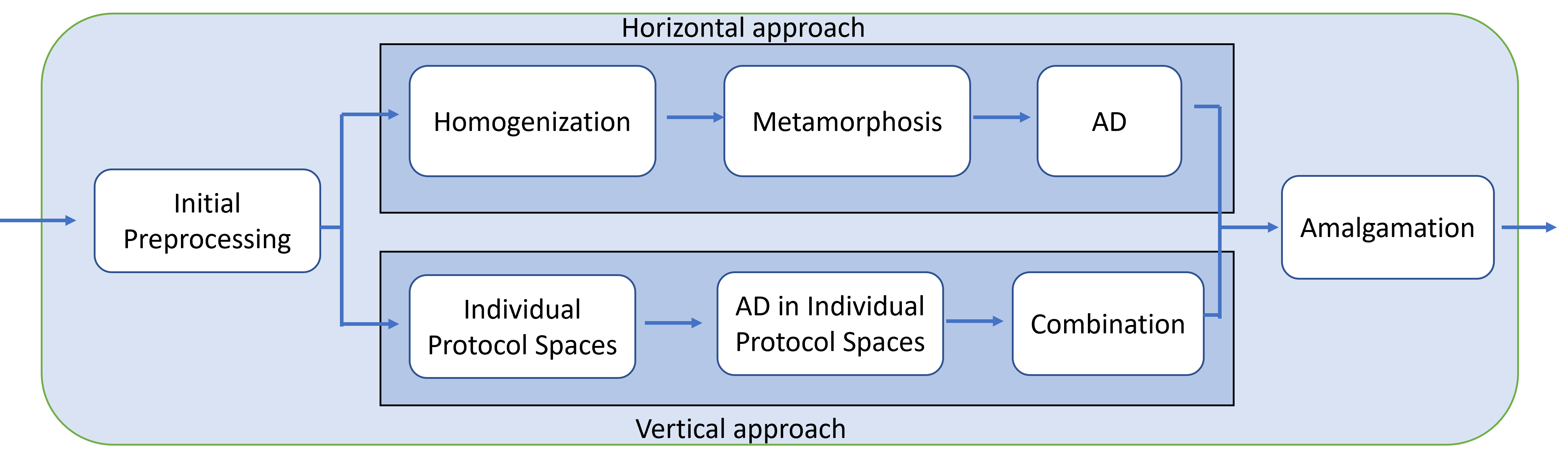} 
     \caption{Honeyboost framework comprising horizontal and vertical approaches. The horizontal approach considers node-level VD time series and identifies anomalous nodes. The vertical approach considers each protocol separately and finds anomalies in each protocol space. }
     \label{fig:framework}
 \end{figure*}

In this paper, we investigate the problem of honeypot aided NAD as a time series analysis problem and present \textit{Honeyboost} -- a hybrid framework to find anomalous nodes. As shown in Figure~\ref{fig:framework}, Honeyboost includes the following approaches:
\begin{enumerate}
\item Horizontal approach: We treat the data from each node as a VD time series and find anomalous time series using a feature based approach, where features are computed from the node-based VD time series.
\item Vertical approach: We focus on each protocol and find anomalous nodes with respect to each protocol using features relevant to that protocol. We then amalgamate the results enabling visual interpretation. 
\end{enumerate} 
Our results demonstrate the following benefits:
\begin{enumerate}
    \item We identify most anomalous nodes before they access the honeypot.
    \item We compare Honeyboost results obtained by the AD method Lookout with a One Class Support Vector Machine (OCSVM) and find the false positive rate of Lookout is much lower compared to the OCSVM.
    \item We rank anomalies, thus enabling prioritization.
    \item We gain deeper insights about the anomalous nodes.
    \item We identify anomalous nodes with suspicious behavior that do not access the honeypot.
\end{enumerate}

The remainder of the paper is organized as follows; In Section~\ref{sec:related} we give a brief introduction to current research in honeypots, anomaly detection and extreme value theory. The datasets for this study, discussed in Section~\ref{sec:exp} are obtained from the Lan-Security Monitoring Project \citep{ochiai2020lan}.  The window-based approach, and the horizontal and vertical methodologies are discussed in Section~\ref{sec:methods1}. We present the early detection results, false positive rates and amalgamation of horizontal and vertical anomalies in Section~\ref{sec:results}. We glean insights from these results in Section~\ref{sec:insights}.  Finally, we present our conclusions in Section~\ref{sec:conclusions}.


\section{Background}\label{sec:related}
In this section we provide the current research related to Honeyboost, including Honeypots, 
Anomaly detection and
Extreme Value Theory (EVT).
 We start with honeypot research. 

\subsection{Honeypots}
Honeypot-based cyber security research can be categorized into two broad areas: \begin{inparaenum} \item honeypot architecture design \citep{Fan2019, Sadasivam2015} and \item threat detection and prevention. \end{inparaenum} These two topics are not mutually exclusive, with some studies encompassing both topics.  For example \cite{Jasek2013}  use Honeyfarms, a centralized collection of honeypots and analysis tools, to analyze and detect advanced persistent threats. In this section we  explore some of the work done on threat detection and prevention using honeypots. 

\cite{Zhan2013} characterize honeypot captured cyber attacks using a statistical framework. They show that these cyber attacks exhibit long-range dependence, i.e. the rate of autocorrelation decay is slower than exponential decay. They use statistical techniques to predict the attack rate, i.e. the number of attacks per time unit. 
\cite{Almohannadi2018} use honeypot log data to evaluate a new threat intelligence technique and find attack patterns. They admit that the large amount of data produced by honeypots is difficult to analyze using general purpose techniques, and instead, use an open search analytics engine called elastisearch.  \cite{La2016} present a game-theoretic model of deception comprising attackers and defenders, using  Bayesian techniques, and verifying their model using simulated data.  As part of the game play, defenders have access to the honeypots. \cite{Moore2015} review production honeypots and analyze data collected from honeypots over a period of time. They explore geographical locations of attacks, IP addresses and ports in this data.  \cite{Shrivastava33} use honeypots to capture attacks on IoT devices and classify them using supervised machine learning algorithms. 

Honeypots are also used to combat targeted attacks. Denial of Service (DoS) attacks can cripple entire networks without finding loopholes in security. 
\cite{Anirudh2017} discuss a honeypot model for mitigating DoS attacks for an Internet of Things (IoT) network. \cite{Tiruvakadu2018} discuss wormhole attacks in the context of mobile, ad-hoc networks and propose a honeypot based solution to confirm these attacks. A wormhole attack consists of two or more attackers strategically placing themselves in a network and creating a tunnel between them. This reduces the shortest path between certain nodes, inducing legitimate traffic to go through the attacker's tunnel. They argue the necessity of an attack confirmation system and use a honeypot to confirm the attacks using a wormhole attack tree. \cite{Zhang2021} use honeypots to validate their adversary detection model in an IoT network. Cybereason, a cyber security technology company, launched a network honeypot in early 2020 to learn the tactics, techniques and procedures of cyber criminals. \cite{Barak2020} discusses the lessons learnt from this experiment and the role of honeypots in critical infrastructure system security. \cite{Handa9514764}  discuss the use of honeypots as a tool to obtain standard cyber defense for small and medium scale businesses. In their book, they describe different cyber security solutions accessible to small enterprises. 

Several recent reviews on honeypots are testament to their rising popularity. \cite{Razali2019} discuss the importance and history of honeypots in information security and review their use in IoT networks. \cite{Seungjin2020} survey honeypot based botnet detection and \cite{Matin2020} review malware detection using honeypots. 

\subsection{Anomaly detection}
As per \cite{hawkins1980identification} \textit{``an anomaly/outlier is an  observation which deviates so much from other observations as to arouse suspicion it was generated by a different mechanism''}. Motivated by this well accepted definition we consider anomalies to be rare observations different from the rest of the points in some feature space. 

Anomaly detection (AD) is an extensively researched topic in many inter-disciplinary research fields. In computer networks and security research, AD  enables detection of unusual patterns in data that may signify new attacks.  While signature based intrusion detection methods reduce the number of false positives, they are ineffective in capturing new attacks. AD methods are commonly used to fill this gap. 

Based on the learning paradigm, anomaly detection methodologies can be broadly categorized into three groups \citep{Goldstein2016}:
\begin{enumerate}
    \item \textit{Supervised anomaly detection}: A model is trained on labeled data, which includes anomalies, and then tested on new data.   
    \item \textit{Semi-supervised anomaly detection}:  Two different definitions exist for semi-supervised anomaly detection. \begin{inparaenum} \item A model is trained on data including both labeled and unlabeled instances \citep{Ruff2020}. Usually the number of labeled data points is much less than the unlabeled ones.   \item A model is trained on data that does not include anomalies \citep{Goldstein2016}.   \end{inparaenum} In both scenarios, the resulting model is tested on new data. 
    \item \textit{Unsupervised anomaly detection}: These techniques do not use a separate set of labeled data to train the model.  
\end{enumerate}
In this study, we focus on unsupervised anomaly detection, as unsupervised methods are better suited to identify new attacks. 

In addition, anomaly detection methods are also categorized as static or dynamic, depending on whether they evolve with streaming data.  The AD literature encompasses a diverse set of  methodologies including density estimation and probabilistic models, distance-based models, one class classification models, reconstruction and deep learning models, and cluster-based and graph-based models. 
For a general survey of AD methodology, see \cite{Wang2019} and for a review of deep learning methodologies refer to \cite{Ruff2020}. 

\subsubsection{Anomaly detection in computer networks}

\begin{figure}[!ht]
     \centering
    \includegraphics[width=0.8\textwidth]{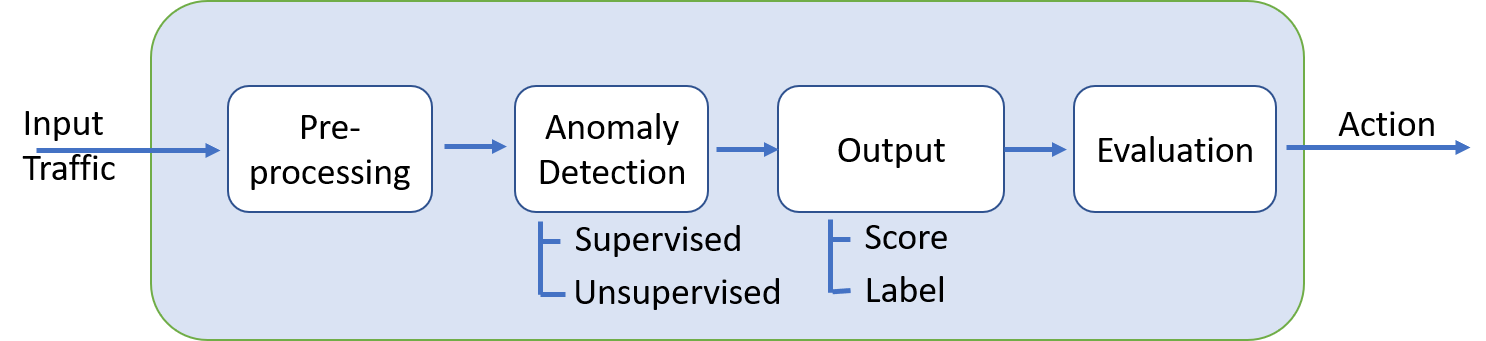}
     \caption{The generic architecture of a NADS as depicted in \cite{AHMED201619}. }
     \label{fig:nads}
 \end{figure}

Figure~\ref{fig:nads} shows the generic architecture of an NADS \citep{AHMED201619}, which includes  pre-processing, anomaly detection, output and evaluation modules.  The preprocessing module  performs feature extraction and feature selection at a packet level \citep{Moustafa2019}. Using either the packet header information alone or packet header along with payload information, a variety of features are computed, then used by the Anomaly Detection (AD) module. The output of the AD module can be real-valued anomaly scores or binary labels signifying whether a data point is an anomaly or not.  The evaluation module is the pre-action phase, where decisions are made based on the anomaly scores/labels. Each module can have several sub-modules feeding into other modules.  

Researchers have used anomaly detection methods to identify attacks and intrusions in computer networks for more than 30 years \citep{Maxion1990}. Furthermore, new anomaly detection methods aimed at specific threats and technologies such as IoT networks are continuously being developed \citep{Naveed2020}. 
A number of comprehensive reviews on network anomaly detection are available \citep{Baddar2014, AHMED201619, Fernandes2019, Moustafa2019}. We briefly discuss the review by \cite{Moustafa2019} and touch upon some latest developments in the area. 

\cite{Moustafa2019} emphasizes that unlike  signature-based methods, anomaly  detection is better suited to capture new attacks. NADS operate by creating a normal profile and identifying deviations from it. In the pre-processing stage, features are extracted from the raw data, then reduced in number by discarding noisy, unimportant features, leaving a smaller set of meaningful features. Next, categorical features are converted to numerical and normalized to enable each feature to contribute equally to anomaly detection. The pre-processing stage is followed by a decision engine, which may have a training phase and a validation and test phase. There are many AD methodologies in NAD, including classification-based methods such as  support vector machines and neural networks, clustering-based methods, deep-learning methods \citep{Sohn2021}, knowledge-based methods, combination-based and statistical methods such as  kernel density estimates and particle filters. The authors discuss these approaches and the available datasets. 

More recent work includes a variety of methods ranging from ensemble methods to fuzzy logic. \cite{Zhou2020} propose an ensemble technique using a modified adaptive boosting method called M-AdaBoost-A to detect intrusions. They extend AdaBoost-A, which can handle class imbalanced data to multi-class classification, and evaluate two variants of the algorithm. \cite{Imrana2021} propose a deep learning approach -- a bi-directional LSTM (long short-term memory) -- for intrusion detection. They evaluate their method using the NSL-KDD dataset. \cite{Hamamoto2018}  propose a method using genetic algorithms and fuzzy logic for network anomalous event detection. Their approach consists of two phases: the  genetic algorithm is used to create a digital signature of the network segment and fuzzy logic is used to identify anomalies. \cite{Khan2021} propose a deep learning method for network anomaly detection using network spectrogram images generated from Fourier transforms. They train a deep convolutional neural network to identify the anomalies. \cite{Liu2021} introduce a NADS using computer log sequences. They discuss the challenges of vectorizing unstructured log messages while preserving semantics in an efficient way. 
\subsection{Extreme Value Theory}
Extreme Value Theory (EVT) is a branch of statistics  used to model rare, extremal events such as 100-year floods and catastrophic financial losses \citep{Reiss2001}. Intuitively, EVT focuses on maxima or minima and represents them using probability distributions. 
EVT is a powerful, yet flexible technique to model different types of extremal behavior. From a statistical point of view, extremes are found in the tail of a distribution. If the tail exhibits exponential decay, that is, if the tail behavior can be written as $\lambda e^{-kx}$ for $\lambda, k  > 0$ where $x$ is the random variable, then that is good news because many methods can be used to model the tail behavior and hence we have a handle on the extremes. For example, the normal distribution has an exponentially decaying tail. The problem arises when we have heavy or fat tails, that is, when the tail behavior exhibits a power law relationship or decays polynomially. When this is the case, the tail behavior can be expressed as  $\lambda x^{-k}$ with $\lambda, k  >0 $. Power law decay is more challenging than exponential decay because it is difficult to control or predict the extremes, meaning there is high probability of getting extreme values. In particular, network traffic is known to have heavy tails \citep{HernandezCampos2004, Ramaswami2014}. Therefore, it is important to use methodologies such as EVT  capable of handling heavy tailed distributions.  

Recent years have seen increasing interest in  EVT based methods for anomaly detection \citep{lookoutPaper,Talagala2020}. We use \textit{Lookout} \citep{lookoutPaper}, an EVT based AD method to detect anomalies in honeypot aided computer networks. Lookout has two main characteristics that are advantageous in a computer network security context: \begin{inparaenum} \item low false positives and a guarantee on false positive probability and \item flexibility to model data without an  assumption on the underlying probability distribution.  \end{inparaenum} 

A low number of false positives is a fundamental characteristic of EVT based AD methods attributed to its theoretical framework. In addition, Lookout gives a guarantee or an upper bound on the false positive probability,  which can be set by the user. Furthermore, Lookout is not limited by any distributional assumptions on the original data.  EVT can handle truncated tails, exponentially decaying tails, polynomially decaying tails  or other types of tail distributions. Depending on the tail of the original distribution, EVT characterizes extremal distributions into 3 types; Gumbel, Frechet and Weibull. These 3 distributions fully describe the space of extreme value distributions, i.e. all extremes fall into one of the above three categories.  A generalized extreme value distribution incorporates these 3 distributions into 1 distribution using an additional shape parameter $\xi$. Lookout estimates this shape parameter using the data, thus choosing the appropriate distribution to model extremes. Therefore, if the data exhibits polynomial decay written as $\lambda x^{-k}$, then Lookout estimates $k$ from the data, which is then used to model extremes and identify anomalies. 


\section{Datasets}\label{sec:exp}

\begin{figure}[!ht]
     \centering
    \includegraphics[scale=0.5]{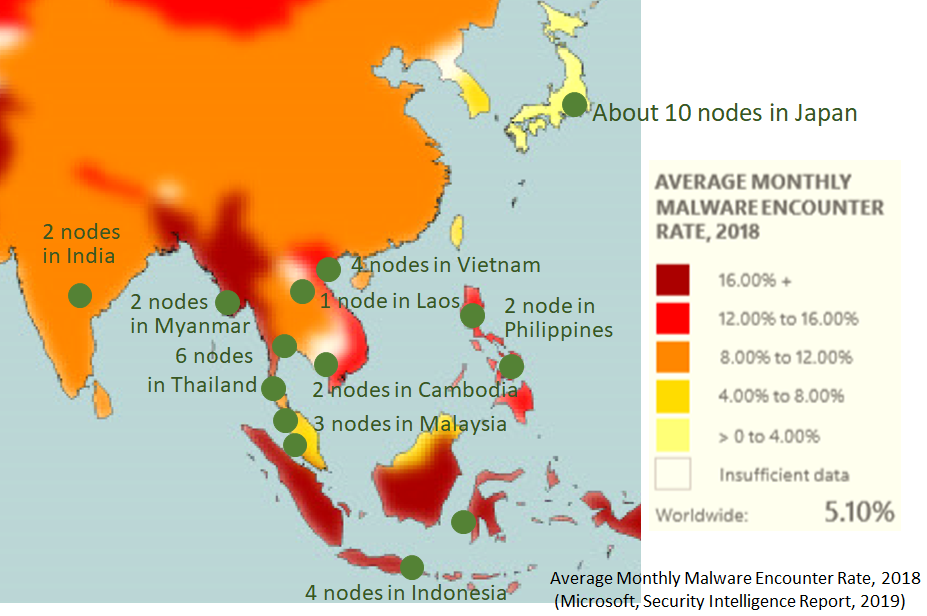}
     \caption{The monitoring device map for the LAN-Security Monitoring Project. The nodes are overlaid on a map taken from the Microsoft security intelligence report \citep{microsoftsir} showing average monthly malware encounter rate of regions.}
     \label{fig:landeploy}
 \end{figure} 

The LAN-Security Monitoring Project \citep{ochiai2020lan} is a research collaboration between 12 ASEAN and SAARC countries led by Japan. The project was deployed in November 2018, and aims to boost cyber-readiness and cyber-resilience among partners through research collaboration especially focusing on the security of local-area networks (LAN). Malware or worms can easily intrude into LANs through phishing-emails or WiFi, even if the network is protected by a firewall or operated under a network address translation (NAT) unit. This project deploys local monitoring devices, honeypots, in LANs because LAN-internal events, such as direct communications between the devices inside the LAN are not visible by the global Internet observatories. Monitoring devices are installed in severe malware-infected countries as shown in Figure~\ref{fig:landeploy}.

\begin{table}[!ht]
	\centering
	\caption{Dataset description -- features by protocol}
	\footnotesize
	\begin{tabular}{p{1cm}p{2cm}p{10cm}}
		\toprule
  Protocol &  Feature & Description  \\
        \midrule
ARP & timestamp & The start of the $n$ second time interval for $n \in \{5, 60, 600 \}$.\\     
    & node address & The MAC address of the source node. \\
    & count & The number of (broadcast) ARP requests made by the node \\
    & degree & The number of IP addresses the node tried to resolve in the time interval of the ARP broadcast request. \\
    \midrule
TCP & timestamp & The start of the 5 second time interval.\\     
    & node address & The MAC address of the source node. \\  
    & num\_ports & The number of ports specified by the TCP packets from this node within the given time interval. \\
    & count & The number of TCP packets the monitoring unit observed from this node. \\
    & avg\_len & The average length of IP packets that deliver the concerned TCP packets \\
    & count\_fin & The number of TCP packets with raised fin\_flag. \\
    & count\_syn & The number of TCP packets with raised syn\_flag. \\ 
    & count\_rst & The number of TCP packets with raised rst\_flag. \\ 
    & count\_psh & The number of TCP packets with raised psh\_flag. \\
    & count\_ack & The number of TCP packets with raised ack\_flag. \\
    & count\_urg & The number of TCP packets with raised urg\_flag. \\
    & count\_ece & The number of TCP packets with raised ece\_flag. \\
    & count\_cwr & The number of TCP packets with raised cwr\_flag. \\
    \midrule
UDP & timestamp & The start of the 5 second time interval.\\     
    & node address & The MAC address of the node. \\  
    & num\_ports & The number of ports specified by the UDP packets from this node within the given time interval. \\
    & count & The number of UDP packets the monitoring unit observed from this node. \\
    & avg\_len & The average length of IP packets that deliver the concerned UDP packets. \\
		 \bottomrule
	\end{tabular}
	\label{tab:firstsetoffeatures}
\end{table}






The datasets we obtained pertained to a single LAN with a honeypot for the time period starting from January 11 2019 until November 15 2020. The honeypot was quiet and did not make any intentional announcements (e.g., advertisements or discovery requests) to the network. However, a suspicious node on the same network segment sometimes directly sends TCP or UDP packets targeting the IP address of the honeypot. Such a node is considered anomalous as there is no legitimate reason to send a TCP/UDP packet to the honeypot. This activity is captured by the software as the packets are directly exchanged with the honeypot. In addition, all broadcast or multicast Ethernet frames including ARP requests, DHCP, NBNS, mDNS, LLMNR packets originating from individual hosts connected to a LAN are also captured by the software. The raw data is then preprocessed to generate features by protocol, which are given in  Table~\ref{tab:firstsetoffeatures}. 

As mentioned above, the datasets we received from the LAN-Security Monitoring Project comprised features by protocol for a specific LAN. As listed in Table~\ref{tab:firstsetoffeatures}, in addition to TCP/UDP data, ARP data was collected at 5s, 60s and 600s time intervals. For each time interval, the node address, count and degree are provided in the dataset. 











\section{Methodology}\label{sec:methods1}

To consider appropriate methodology, we take into account that the datasets span  22 months. Using all the data in one batch and detecting anomalous nodes would amount to detecting anomalies at the end of the 22 month period. Instead, we use a more continuous approach to detect anomalies using a window-based model. A window-based approach uses the data in the time window to detect anomalies. That is, it does not see data outside the window. Both horizontal and vertical approaches use window-based models for anomaly detection. We combine the anomalies found in the amalgamation stage.

\subsection{Sliding and expanding time window models}
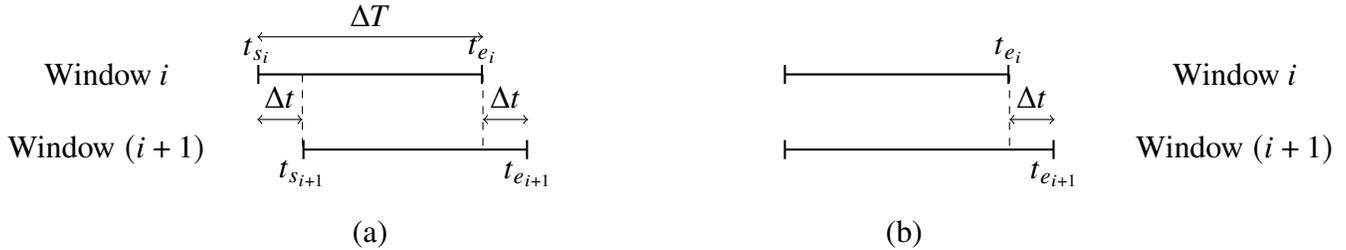
\begin{figure}[!ht]
     \centering
     \begin{tikzpicture}[scale=2]
      \draw[thick,|-|] (0,0) -- (1.5,0) node[anchor=south]{$t_{e_i}$};
      \node[anchor=south] at (0,0) {$t_{s_i}$};
      \draw[thin,<->] (0,0.25) -- (1.5,0.25) node[midway, anchor=south]{$\Delta T$};
      \draw[thick,|-|] (0.3,-0.5) -- (1.8,-0.5) node[anchor=north]{$t_{e_{i+1}}$};
      \node[anchor=north] at (0.3,-0.5) {$t_{s_{i+1}}$};
      \draw[thin,<->] (0,-0.3) -- (0.3,-0.3) node[midway, anchor=south]{$\Delta t$};
      \draw[thin,<->] (1.5,-0.3) -- (1.8,-0.3) node[midway, anchor=south]{$\Delta t$};
      \draw [dashed] (0.3,0) -- (0.3, -0.5);
      \draw [dashed] (1.5,0) -- (1.5, -0.5);
      \node at (-1,0) {Window $i$};
      \node at (-1,-0.5) {Window $(i+1)$};
      \node[anchor=north]  at (0.75,-0.9) {(a)};
      
      \draw[thick,|-|] (3.5,0) -- (5,0) node[anchor=south]{$t_{e_i}$};
      \draw[thick,|-|] (3.5,-0.5) -- (5.3,-0.5) node[anchor=north]{$t_{e_{i+1}}$};
      \draw[thin,<->] (5,-0.3) -- (5.3,-0.3) node[midway, anchor=south]{$\Delta t$};
      \draw [dashed] (5,0) -- (5.0, -0.5);
      \node[anchor=north]  at (4.3,-0.9) {(b)};
      \node at (6.5,0) {Window $i$};
      \node at (6.5,-0.5) {Window $(i+1)$};
    \end{tikzpicture}
\caption{Sliding and expanding windows: (a) The sliding window model. This is used in the horizontal approach and in the vertical approach for ARP. (b) The expanding window model. This is only used for TCP and UDP in the vertical approach.  }
\label{fig:windows}
\end{figure} 

 Figure~\ref{fig:windows} shows the different window models used for horizontal and vertical approaches. For the horizontal approach, we use a sliding window model for all three protocols. We denote the window width by $\Delta T$ and the step size by $\Delta t$. Suppose the $i^{\text{th}}$ time window starts at $t_{s_i}$ and ends at $t_{e_i}$. Then $t_{e_i} = t_{s_i} + \Delta T$.  The  $(i+1)^{\text{st}}$ time window would start at  $t_{s_{i+1}} = t_{s_i} +\Delta t$ and ends at $t_{e_{i+1}} = t_{s_{i+1}} + \Delta T = t_{e_{i}} + \Delta t$. This is illustrated in Figure~\ref{fig:windows}(a). Thus, the horizontal approach considers data in the time window $\left[t_{s_i}\, , t_{e_i}  \right]$ for each window $i$.

The vertical approach is somewhat different because it considers each protocol separately to detect anomalies.  We use two different window models for the vertical approach, based on the protocol. For ARP, we use the same sliding windows as in the horizontal approach. For TCP and UDP we use an expanding window model shown in Figure~\ref{fig:windows}(b). 
The honeypot captures all the broadcast ARP requests, but only captures TCP or UDP packets directed at it. Consequently, the number of TCP and UDP packets, compared to ARP packets, are low. Thus, a sliding window model is not suitable for TCP and UDP separately, for the vertical approach. Thus, for each ARP window ending at time $t_{e_i}$, we use TCP and UDP data for $t \leq t_{e_i}$.

While there is no definitive guide to choosing the window size $\Delta T$ or the step size $\Delta t$, often application specific considerations and   trial and error are used in selecting these parameters. For each window $i$, that is, at every $t_{e_i},$ Honeyboost outputs a list of anomalous nodes identified from the data in that window. As such, the step size $\Delta t$ is the frequency of Honeyboost output, which can selected according to requirement. We have used $\Delta t = 60 \times 60$ seconds in this experimental study, i.e., we have configured Honeyboost to output anomalous nodes every hour. 

In contrast, the window size $\Delta T$ determines the amount of data the algorithm looks at to detect outliers. Smaller window sizes give rise to higher fluctuations. For example, using a window size of 1 hour results in the volume of data being high at peak activity hours and much lower at 4 am in the morning. This is known as seasonality of the data. Often with network traffic data, there could be multiple seasonal patterns ranging from hourly patterns to  daily patterns. Similar to the hourly fluctuations, there is lower traffic on some days of the week compared to other days. Therefore, it is important to choose a window size  not affected by such seasonal patterns. In addition, the higher fluctuations in data volume from smaller window sizes give rise to higher false positives. Here, to mitigate seasonal fluctuations and to keep the false positives low, we use $\Delta T = 1 \, \text{week} = 7 \times 24 \times 60 \times 60 $ seconds, and configure Honeyboost to predict anomalous nodes every hour from one week's worth of data ending at that hour.



\subsection{Horizontal approach}

\begin{figure}[!ht]
     \centering
    \includegraphics[scale=0.5]{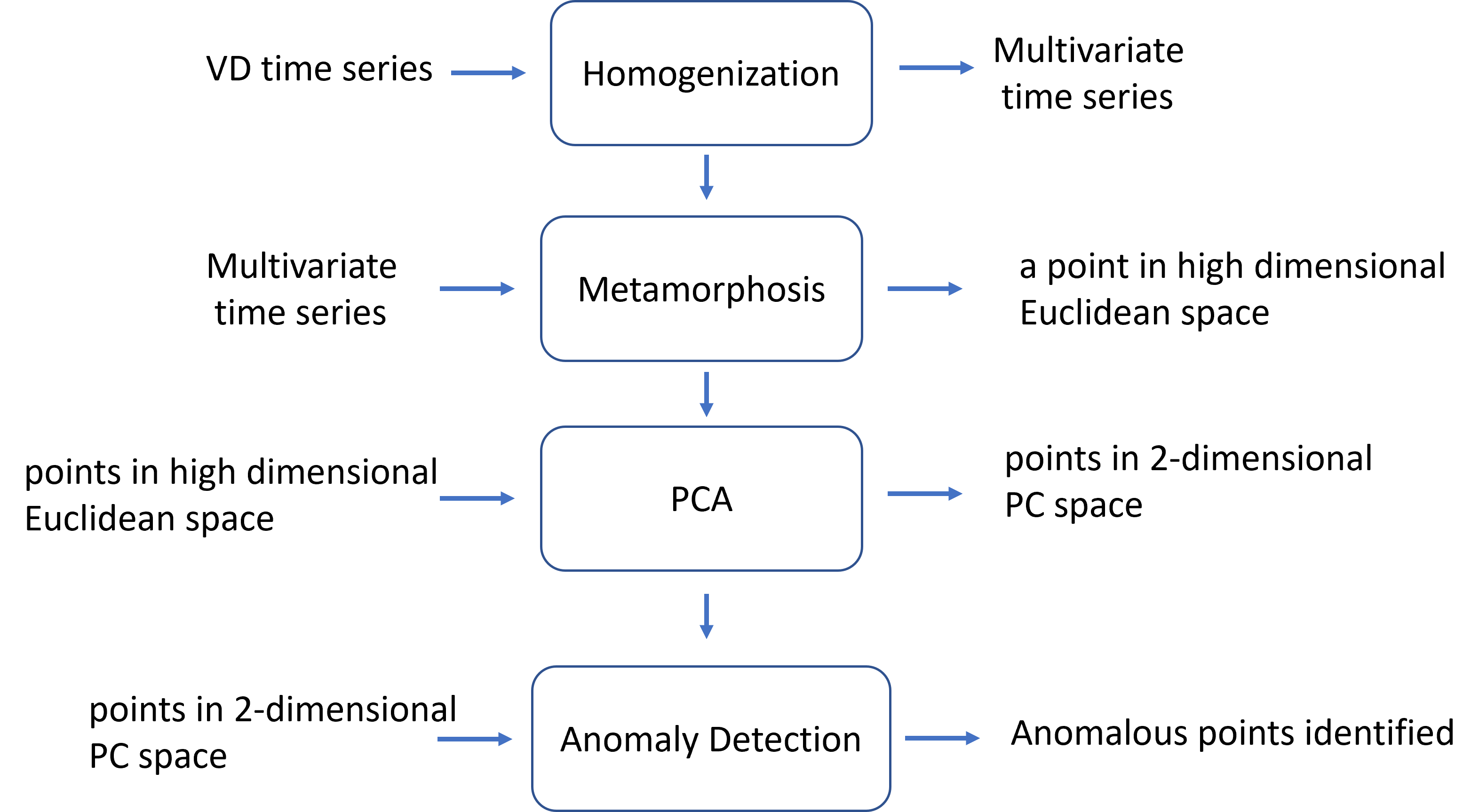}
     \caption{The input and output data type of each process in the horizontal approach.}
     \label{fig:horizontalapproach}
 \end{figure} 
 
Using the sliding window model, the pre-processed features  for all three protocols, by node, are sorted by time, resulting in, for each time window and each node, a set of varying dimensional (VD) time series as depicted in Figure~\ref{fig:horizontalonenode}. Figure~\ref{fig:horizontalapproach} illustrates the processes involved in this approach, and the input and the output data type for each process.

Firstly, the VD time series is reshaped to a multivariate time series using homogenization (see Table~\ref{tab:homogenizedExamples}). This essential step results in transporting the research problem to a space where we can use existing time series analysis methodology. However, finding anomalous time series from a collection of multivariate time series is a challenging task. Hence, we transform each multivariate time series to a point in high dimensional space using  metamorphosis (Figure~\ref{fig:metamorphosis}), a feature-based time series approach popular in time series forecasting. This step transforms the problem of finding anomalous multivariate time series to finding anomalous points in Euclidean space, an easier problem. However, anomaly detection in high dimensional data is prone to error due to the curse of dimensionality. Therefore, the dimensions of this data are reduced using Principal Component Analysis (PCA) and  only the first 2 PC scores are kept. Anomalies are then detected in this 2-dimensional PC space  using Lookout and a One Class Support Vector Machine (OCSVM). Lookout is an Extreme Value Theory based anomaly detection method and achieves lower false positives compared to the OCSVM. These processes are detailed below. 


\subsubsection{Homogenization: \textit{constructing a multivariate time series from a VD time series}} \label{subsec:homogenization}  

A multivariate time series is constructed using the VD time series for each node.  This transformation, called  \textit{homogenization}, is shown in Table~\ref{tab:homogenizedExamples} via an example. The resulting multivariate time series contains 6 numeric features characterizing  important aspects of each protocol along with the node address, timestamp and  protocol name.    

As seen in Table~\ref{tab:firstsetoffeatures} the dataset from each protocol contains the node address, the timestamp and some other numerical features for each observation. These numerical features are depicted in parenthesis in the example VD time series given in Table~\ref{tab:homogenizedExamples}. In order to give each protocol the same importance, we represent the numeric features of each protocol by two key features. The two numeric features of ARP (see Table~\ref{tab:firstsetoffeatures}) namely degree and count are included as part of the 6 features in the multivariate time series.    

TCP, on the other hand, has 11 numeric attributes per node at each timestamp (see Table~\ref{tab:firstsetoffeatures}). We reduce the dimension of this 11-dimensional feature space to 2, by using Principal Component Analysis (PCA). The first 2 principal component scores PC1 and PC2 for TCP then become part of the 6 features in the multivariate time series. Similarly UDP has 3 numeric attributes at each timestamp; PCA is performed for the UDP space and the first 2 PC scores used.  

Thus, each node is homogenized to have the following 8 features: \begin{inparaenum}   \item timestamp; \item protocol name; \item ARP degree; \item ARP count; \item TCP PC1; \item TCP PC2; \item UDP PC1; and \item UDP PC2. \end{inparaenum} Thus a VD time series for a given node with observations at $k$ different time points is transformed into a $k \times 8$ matrix denoting an 8 dimensional time series with $k$ time points. The node address is a grouping attribute that does not participate in the statistical analysis. That is, once the anomalous points are identified, the node address is only used to recognize the responsible node. 



\begin{table}[!ht]
\caption{Homogenization: An example VD time series (left) reshaped to a multivariate time series (right)}
\parbox{.35\linewidth}{
\centering
\begin{tabular}{llll}
		\toprule
    \rot{Node Address} & \rot{Timestamp} &  \rot{Protocol} & Features \\
        \midrule
    N1 &  30 &  ARP & $(10, 12)$  \\
    N1 &  55 &  TCP &  $(80, 2, 6, 0, 2, 0, 0, 0, 0, 1, 1)$ \\
    N1 &  85 &   UDP &  $(3702, 2, 652)$ \\
		 \bottomrule
	\end{tabular}
}
\hfill
\hfill
$\Longrightarrow$
\parbox{.5\linewidth}{
\centering
	\begin{tabular}{lllllllll}
		\toprule
    \rot{Node Address} & \rot{Timestamp} &  \rot{Protocol} &  \rot{ARP Count} & \rot{ARP Degree} & \rot{TCP PC1} & \rot{TCP PC2} & \rot{UDP PC 1} & \rot{UDP PC2} \\
        \midrule
      N1 & 30 &  ARP &  10 & 12 & 0 & 0 & 0 & 0 \\
      N1 & 55 &  TCP &  0 & 0 & 2.1 & 1.7  & 0 & 0 \\
      N1 & 85 &  UDP & 0 & 0 & 0 & 0 & 3.6 & 0.4 \\
		 \bottomrule
	\end{tabular}
}
\label{tab:homogenizedExamples}
\end{table}

\subsubsection{Metamorphosis: \textit{Mapping a time series to a point in high-dimensional space}} \label{sec:metam} 

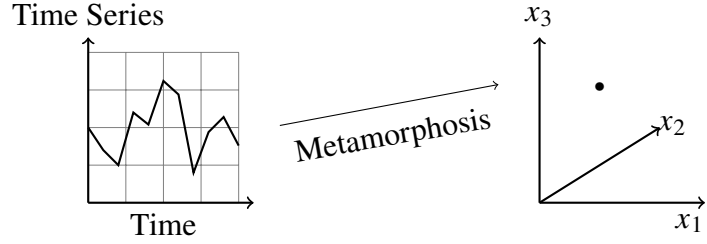
\begin{figure}[!ht]
     \centering
     \begin{tikzpicture}[scale=2]
        \draw[step=0.25cm, gray,very thin] (0,0) grid (1,1);
        \node[anchor=north]  at (0.5,0) {Time};
        \draw[thick,->] (0,0) -- (1.1,0); 
        \draw[thick,->] (0,0) -- (0,1.1) node[anchor=south]{Time Series};
        \draw[thick] (0, 0.5) -- (0.1,0.35) -- (0.2,0.25) -- (0.3,0.6) -- (0.4,0.52) -- (0.5,0.81) -- (0.6,0.72) -- (0.7,0.2) -- (0.8,0.47) -- (0.9,0.57) -- (1,0.38);

        \node[anchor=north] at (4,0) {$x_1$};
        \node[anchor=north east] at (4.05,0.65) {$x_2$};
        \node[anchor=south] at (3,1.1) {$x_3$};
        \draw[thick,->] (3,0) -- (4.1,0) ; 
        \draw[thick,->] (3,0) -- (3,1.1) ;
        \draw[thick,->](3,0)--(3.8,0.5);
        \node at (3.4, 0.75) {\textbullet};
        
        \node (node1) at (1.2, 0.5) {};
        \node (node2) at (2.8, 0.8) {};
        \draw[->] (node1) -- (node2);
         \node[anchor=north, rotate=10] (node3) at (2, 0.6) {Metamorphosis};
    \end{tikzpicture}
\caption{A simpler version of metamorphosis: The figure on the left shows a univariate time series, which gets mapped to a point in $\mathbb{R}^3$ on the right. In reality, we consider a multivariate time series for each node, and map it to a point in $\mathbb{R}^{17}$. }
\label{fig:metamorphosis}
\end{figure} 

For each time window $i$, where $t \in \left(t_{s_i}, t_{e_{i}} \right)$, and each node, we consider a multivariate time series. This gives us $m$ multivariate time series if there is traffic from $m$ nodes present in that time window.  We then compute features and transform this time series into a point in a high-dimensional space. As a time series gets transformed into a point, we call this process \textit{metamorphosis}. Figure~\ref{fig:metamorphosis} illustrates this process. There is a bijective mapping between the nodes and the points in the high-dimensional space, wherein each node is uniquely identified by a point in the high dimensional space and each point in the high-dimensional space denotes the behavior of a specific node in that time window. \\

\begin{figure}[!ht]
     \centering
     \begin{tikzpicture}[scale=2.5]
        \draw[step=0.25cm, gray,very thin] (0,0) grid (1,1);
        \node[anchor=north]  at (1,0) {ARP count};
        \draw[thick,->] (0,0) -- (1.1,0); 
        \draw[thick,->] (0,0) -- (0,1.1) node[anchor=south]{ARP degree};
        \draw[thick] (0.95,0.8) -- (0.1,0.35) -- (0.95,0.38) --(0.3,0.62) --(0.7,0.2) -- (0.65, 0.9) --(0.15, 0.2) ;
        \node[anchor=south] at (0.95,0.8) {$t_1$};
        \node[anchor=south] at (0.1,0.35) {$t_2$};
        \node[anchor=south] at (0.95,0.38) {$t_3$};
        \node[anchor=south] at (0.3,0.62) {$t_4$};
        \node[anchor=north] at (0.7,0.2) {$t_5$};
        \node[anchor=south] at (0.65, 0.9) {$t_6$};
        \node[anchor=north] at (0.15, 0.2) {$t_8$};
        \node[anchor=north]  at (0.5,-0.20) {(a)};
        
         
        \draw[step=0.25cm, gray,very thin] (2,0) grid (3,1);
        \node[anchor=north]  at (3,0) {TCP PC1};
        \draw[thick,->] (2,0) -- (3.1,0); 
        \draw[thick,->] (2,0) -- (2,1.1) node[anchor=south]{TCP PC2};
        \draw[thick] (2.1, 0.5) -- (3,0.38) --(2.5,0.9) -- (2.55, 0.2);
        \node[anchor=south] at (2.1, 0.5) {$t_2$};
        \node[anchor=south] at (3,0.38) {$t_4$};
        \node[anchor=south] at (2.5,0.9) {$t_5$};
        \node[anchor=north] at (2.55, 0.2) {$t_9$};
         \node[anchor=north]  at (2.5,-0.20) {(b)};
        
        
         \draw[step=0.25cm, gray,very thin] (4.0,0) grid (5.0,1);
         \node[anchor=north]  at (4.5,0) {ARP count};
        \draw[thick,->] (4.0,0) -- (5.1,0); 
        \draw[thick,->] (4.0,0) -- (4.0,1.1) node[anchor=south]{ARP degree};
        \filldraw[black] (4.95,0.8) circle (0.5pt) node[anchor=south] {$t_1$};
        \filldraw[black] (4.1,0.35)  circle (0.5pt) node[anchor=south] {$t_2$};
        \filldraw[black] (4.95,0.38)  circle (0.5pt) node[anchor=south] {$t_3$};
        \filldraw[black] (4.3,0.62)  circle (0.5pt) node[anchor=south] {$t_4$};
        \filldraw[black] (4.7,0.2)  circle (0.5pt) node[anchor=south] {$t_5$};
        \filldraw[black] (4.65, 0.9)  circle (0.5pt) node[anchor=south] {$t_6$};
        \filldraw[black] (4.15, 0.2)  circle (0.5pt) node[anchor=north] {$t_8$};
        \draw[thick,-] (4,0.34) -- (5, 0.62); 
          \node[anchor=north]  at (4.5,-0.20) {(c)};
        
   \end{tikzpicture}
     \caption{(a) The ARP degree and count for a hypothetical node at different time points, with successive time points connected by line segments. The total length of these line segments is a feature. (b) The TCP PC1 and PC2 coordinates for the same node, some of which occur at different time points; the total length of these line segments is calculated. (c) The line of best fit for points in the ARP space. The slope, intercept and the sum of squares errors are calculated using this line of best fit. }
     \label{fig:arpandtcpspace}
 \end{figure}
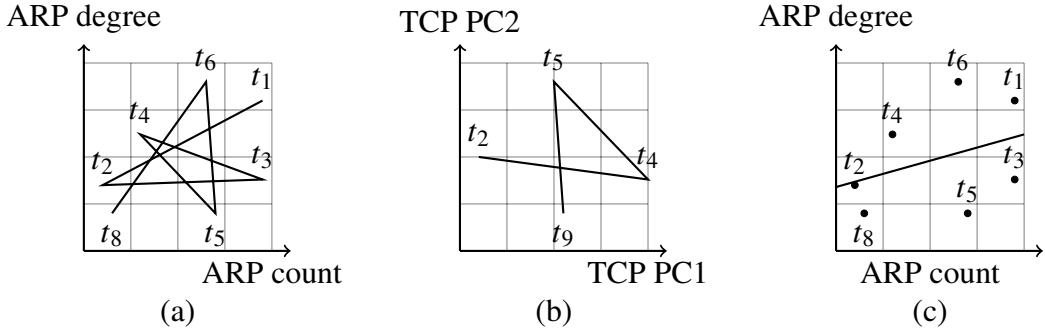

\noindent
\textbf{The features} \\
For any given node $N_j$ the following features  are computed from the multivariate time series associated with $N_j$:  
\begin{enumerate}  
\item The maximum time difference for the time series $\max t - \min t$ for $t \in \left(t_{s_i}, t_{e_{i}} \right)$ for node  $N_j$
\item The number of protocols used by $N_j$. 
\item The number of TCP calls made by $N_j$. 
\item The number of UDP calls made by $N_j$. 
\item The total length of the line segments in $\mathbb{R}^6$ using the 6 protocol features, ARP count, ARP degree, TCP PC1, TCP PC2, UDP PC1 and UDP PC2: Suppose $\{ \bm{x}_{t_k} \}_{k=1}^\ell$ denotes the time series for $N_j$ in $\mathbb{R}^6$. The standard Euclidean norm gives the length of the line segments, viz.,  $ \sum_{k=1}^{\ell-1}\lVert \bm{x}_{t_{k+1}} - \bm{x}_{t_k}  \rVert$. In the example given in Table~\ref{tab:homogenizedExamples} the total length of the line segment in $R^6$ would be  $\lVert(10, 12, 0, 0, 0, 0) - (0,0,2.1, 1.7, 0, 0) \rVert + \lVert (0,0,2.1, 1.7, 0, 0) - (0,0,0,0,3.6,0.4) \rVert $.

\item The total length of the line segments in the ARP space spanned by ARP count and ARP degree: While similar to the previous feature this only takes the length of the line segments using the ARP count and ARP degree. Figure~\ref{fig:arpandtcpspace}(a) illustrates this ARP space for a hypothetical node. At times $t_1$, $t_2$, $t_3$, $t_4$, $t_5$, $t_6$  and $t_8$ this node has made ARP calls. The total length of the line segments is given by $\sum_{t, t' \in S} \lVert \bm{x}_{t'} - \bm{x}_{t}  \rVert$, where $S =\{t_1,t_2,t_3,t_4,t_5,t_6,t_8\}$ and $t$ and $t'$ denote successive time points in $S$.
\item Next we fit a line to the points in the ARP space using linear regression. We denote the points in the ARP space by $\{ \bm{x}_{t_k} \}_{k=1}^\ell$ where $\bm{x}_{t_k} = \left(x_{kc}, x_{kd}\right)$ where $x_{kc}$ denotes the count and $x_{kd}$ denotes the degree at time $t_k$.  Then the line of best fit is given by the equation

\begin{equation}\label{lineofbestfit}
    \hat{x}_{kd} = m {x}_{kc} + c \, , 
\end{equation}
where $m$ and $c$ denotes the slope and intercept, which are given by
\begin{align}\label{slope}
   m  &= \frac{\ell \sum_k {x}_{kc}{x}_{kd}  - \sum_k {x}_{kc}  \sum_k {x}_{kd} }{\ell \sum_k {x}_{kc}^2 - \left( \sum_k {x}_{kc}\right)^2 } \, ,  \\
\text{and} \qquad c & = \frac{1}{\ell} \left( \sum_k  {x}_{kd} - m  \sum_k {x}_{kc} \right)   \, . \end{align}
The quantity $\hat{x}_{kd}$ is the predicted ARP degree by this linear model. The sum of squares errors is given by
\begin{equation}\label{sse}
 \text{SSE} = \sum_{k}\left(\hat{x}_{kd} - {x}_{kd}  \right)^2 \, , 
\end{equation}
that is the squared difference between the actual and the predicted degree values summed over the number of ARP calls. A low value of SSE indicates that the line is a better fit for the data compared to a high SSE value. These three features, slope, intercept and SSE,  are added to the feature pool. An example line of best fit is shown in Figure~\ref{fig:arpandtcpspace}(c). 
\item The equivalent of features 6 and 7 are computed for TCP. This gives the total length of the line segments in the TCP PC1, PC2 space along with the slope, intercept and sum of errors squares of the line of best fit (see Figure\ref{fig:arpandtcpspace}(b)). 
\item Similarly, the length of the line segments in the UDP PC1, PC2 space and the slope, intercept and SSE of the line of best fit are computed. 
\end{enumerate}
We now have 17 features describing the multivariate time series for each node, thus transforming each time series to a point in $\mathbb{R}^{17}$. Therefore, a time window containing data from $m$ distinct nodes gives rise to $m$  time series, which get transformed to $m$ points in $\mathbb{R}^{17}$, with each point denoting the behavior of a node in that time window. \\

\subsubsection{Dimension reduction and anomaly detection: \textit{Identifying anomalies after dimension reduction} }\label{sec:refanomalies}
In higher dimensions it is hard to differentiate anomalies from non-anomalies because points are far away from each other. Therefore, we reduce dimensions using PCA and use the first 2 dimensions in the PC space for anomaly detection.  We use the algorithm \textit{Lookout} \citep{lookoutPaper} and an OCSVM to identify anomalies in this two dimensional space. Lookout uses Extreme Value Theory (EVT) \citep{coles2001introduction} and leave-one-out kernel density estimates to identify anomalies. We briefly describe the algorithm \textit{Lookout} next. 
\begin{enumerate}
    \item To compute kernel density estimates, a bandwidth parameter is required. However, the bandwidth suited for representing the data in general is not suitable for anomaly detection. Using persistent homology \citep{ghrist2008barcodes}, a methodology in topological data analysis, Lookout chooses a bandwidth appropriate for anomaly detection.
    \item Using this bandwidth, kernel density estimates (kde) and leave-one-out kernel density estimates are computed for the data.
    \item As anomalies are rare, they lie in low density regions in a feature space. The bandwidth selected in step 2 results in anomalies having low kde and leave-one-out kde values. 
    However, we want to identify the anomalies, not just give an anomaly score for all the data points. A threshold is needed so that anomalies with kde values lower than this threshold are declared anomalous. We set a threshold by computing the probabilities of the points using EVT. Specifically, by using Generalized Pareto Distributions. 
    \item If the probability of a point is below a predefined threshold $\alpha$, it is declared anomalous. 
\end{enumerate}
For this work we set the threshold as $\alpha = 0.1$, i.e., if the probability $p$ of a point being present using the Generalized Pareto Distribution is less than 0.1, we identify it as an anomaly. Using this probability $p$ the anomaly score is given by 
\begin{equation}
    s = (0.1 - p) \times 100\, , 
\end{equation}
resulting in the anomaly scores lying between $0$ and $10$ with data points with  zero probability having an anomaly score of 10. See \cite{lookoutPaper} for more details on Lookout. 

Detecting a stream of false anomalies decreases  confidence in the NAD system making a technique with low false positives  attractive in this domain. EVT related methods only identify extremes as anomalies. As  such, we expect Lookout to have a low false positive rate.  One-class SVMs (OCSVM) are often used for anomaly detection in an unsupervised setting.  Given the challenges posed by high false positives in NAD, we compare the performance of Lookout with an OCSVM. 


As the sliding time windows overlap, some nodes are identified as anomalies in multiple time windows, while others appear as anomalous only in one time window. If a node is identified as anomalous in multiple time windows, it is a persistent anomaly, and provides grounds for investigation. 

\subsection{Vertical Approach}\label{sec:methods2}
The horizontal approach investigates the activity of each node as a time series and finds anomalous time series. In contrast, the vertical approach focuses on each protocol to find anomalous activity.  All observations in a single protocol have the same number of attributes, and thus the same dimension, making the vertical approach much simpler as it does not face the challenge of assembling  varying dimensional data into a single construct.  The vertical approach comprises the following steps:
\begin{enumerate}
    \item The numerical feature space for each protocol has different dimensions, with ARP being 2-dimensional feature space, TCP  11-dimensional and UDP  3-dimensional (see Table~\ref{tab:firstsetoffeatures}). Identifying anomalies in high dimensions is difficult, and we consider a 2-dimensional space for each protocol. For ARP we take the count and the degree. For TCP and UDP we perform PCA and take the first two PC scores.
    \item Lookout and an OCSVM are used to find anomalies in each protocol's 2D space. 
    \item The anomalies are combined by node and time window to gain better insights. 
\end{enumerate}

For the vertical approach, each point in a protocol 2D space represents a single communication from a node using that protocol. Nodes communicating multiple times in a time window are represented by multiple points in the vertical approach. Thus, if a node $N$ communicates using ARP, TCP and UDP $m_1$, $m_2$ and $m_3$ times respectively, it will result in $m_1$ points in the ARP space, $m_2$ points in the TCP space and $m_3$ points in the UDP space.

Finally, the anomalies identified from the horizontal and vertical approaches for each time window are amalgamated to get a better understanding of the behavior of the nodes. A node identified as anomalous by both horizontal and vertical approaches gains higher priority for further inspection than a node identified by a single approach. 

\section{Results}\label{sec:results}

This section details the results found from using Honeyboost to identify anomalous nodes. We start with comparing the two AD methods used in Honeyboost with regards to early detection of anomalous nodes.

\subsection{Early detection}
 The two AD methods in Honeyboost -- Lookout and OCSVM -- perform similarly and identify most nodes as anomalous before they access the honeypot. Table~\ref{tab:earlyanomalies} shows the anomalous nodes, the earliest time they access the honeypot and the results of the 2 AD methods. The Table gives the earliest time each AD method identifies a node as anomalous. If this detection happens  before the node accesses the honeypot, it is labeled as early detection, denoted by \textit{Early} under the heading \textit{Status}. The table also indicates the time each node is identified as anomalous under the heading \textit{Time Difference} with positive values indicating early detection. As the window step size is 3600 seconds, any time difference in the interval $[-3600, 0)$ indicates  the anomaly was identified in the same time window.  

Of the 15 nodes that access the honeypot, Lookout identifies 13 nodes before they access the honeypot and 2 nodes (N220 and N225) in the same time window that they access the honeypot. The OCSVM identifies all nodes before they access the honeypot. 



\begin{table}[!ht]
	\centering
	\caption{Early detection of honeypot anomalies - amalgamated results using the horizontal and vertical approaches (Early = Early Detection, SW = Same Window)}
	\footnotesize
	\begin{tabular}{cp{2cm}p{2.5cm}p{1cm}p{2.5cm}p{0.1cm}p{2.5cm}p{1cm}p{2.5cm}}
		\toprule
    \multirow{3}{*}{Node}  & \multirow{3}*{\begin{tabular}{l} Earliest \\ Honeypot \\ Time  \end{tabular} } &  \multicolumn{3}{c}{Lookout Results} & & \multicolumn{3}{c}{OCSVM Results} \\
   \cmidrule{3-5}  \cmidrule{7-9} & & Earliest Anomaly Detection Time & Status  &  Time Difference (Positive = Early) & & Earliest Anomaly Detection Time & Status  &  Time Difference (Positive = Early) \\
        \midrule
N021  &  1598842055   &         1550034180 & Early  &  48807875 & & 1547269200 & Early & 51572855\\
N039   &  1553585825   &         1553585820 & Early &         5 & & 1547449140 & Early & 6136685\\
N046   &  1553751195   &         1549530120 & Early &   4221075 & & 1547521980 & Early & 6229215 \\
N132   &  1566462345   &         1561276560 & Early &   5185785 & & 1554347700 & Early & 12114645\\
N135   &  1554351595   &         1554351540 & Early &        55 & & 1554351540 & Early &   55\\
N153   &  1563528080   &         1555997940 & Early &   7530140 & & 1554803160 & Early & 8724920\\
N155   &  1565313000   &         1557801000 & Early &   7512000 & & 1554952440 & Early &  10360560\\
N158   &  1555312055   &         1555311960 & Early &        95 & & 1555311900 & Early &      155\\ 
N159   &  1555313055   &         1555312740 & Early &       315 & & 1555312740 & Early &     315\\
N163   &  1567236235   &         1558260780 & Early &   8975455 & & 1557228000 & Early &  10008235\\ 
N171   &  1565059350   &         1563260880 & Early &   1798470 & & 1558065360 & Early &   6993990\\
N219   &  1564535840   &         1564535760 & Early &        80 & & 1564535760 & Early &       80\\ 
N220   &  1565758430   &         1565758435 & SW    &  -5       & & 1564552620 & Early &     1205810\\
N225   &  1565758055   &         1565758080 & SW    &  -25      & & 1565695320 & Early &     62735\\
N239   &  1575507740   &         1572234480 & Early &   3273260 & & 1568005320 & Early &    7502420\\
		 \bottomrule
	\end{tabular}
	\label{tab:earlyanomalies}
\end{table}

Given that Honeyboost identifies anomalies before they access the honeypot, it is important to consider the number of false positives it generates. Note that if a node is tagged as anomalous the first time it communicates with another node, Honeyboost would detect all anomalies early. However, such a strategy would create a barrage of false positives,  greatly reducing the confidence in the system. For example, a home security system that constantly trips the alarm. Therefore, it is imperative to consider the false positive rate, i.e., early detection is valuable only if the false positive rate is low.  

\subsection{False positives}
\begin{figure}[!ht]
     \centering
    \includegraphics[scale=0.9]{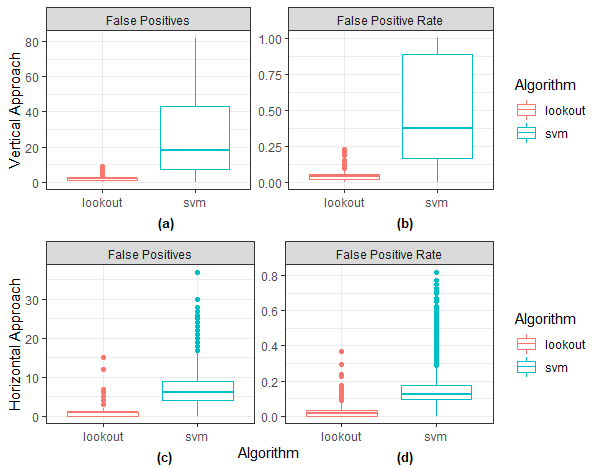}
     \caption{False positives using Lookout and the OCSVM for all time windows. Figures (a) and (c) give the number of false positives for the vertical and horizontal approaches. Figures (b) and (d) give the false positive rates for the two approaches. }
     \label{fig:horizontalinandout}
 \end{figure} 
 
We investigate the number of false positives and the false positive rate,  defined as the ratio of the number of false positives to the total number of negatives for both the horizontal and the vertical approaches. Figure~\ref{fig:horizontalinandout} shows the false positives and the false positive rates for both Lookout and OCSVM for both approaches. Figures \ref{fig:horizontalinandout}(a) and (c) show boxplots of the number of false positives for each time window. The false positive rate is given in Figures \ref{fig:horizontalinandout}(b) and (d). 

Lookout achieves a low false positive rate compared to the OCSVM for both approaches.  In the vertical approach, the OCSVM identifies many nodes as anomalies in each window with a median of 20 false positives. In some windows the OCSVM outputs 80 false positives. This large number of false positives causes the false positive rate to have a median of 37\% and an upper quartile of 87\%. The highest false positive rate for the OCSVM for the vertical approach is close to 1. 

It is interesting to note that the vertical approach produces more false positives for the OCSVM compared to the horizontal approach. This is due to the number of data points in a time window for each approach. Suppose a given time window has $N$ ARP calls belonging to $n$ nodes with $N >>n$. Generally $N$ is orders of magnitude higher than $n$. The horizontal approach constructs a data point for each node in each window giving rise to $n$ data points. In contrast, the vertical approach has $N$ data points. Thus, it is clear the OCSVM identifies more nodes as anomalous when there are a large number of data points. 

Given that Lookout identifies anomalies early and still has a low false positive rate, we focus on the results produced by Lookout for the remainder of the paper. 

\subsection{Amalgamating horizontal and vertical anomalies}

For each window, the anomalies detected by  the horizontal and vertical approaches are amalgamated. Table~\ref{tab:amalgamation} shows the anomalies identified and amalgamated for the vertical and horizontal approaches for each of 4 windows, using Lookout.  In window 2270,  2 anomalies were identified: N046 is identified by both the horizontal and vertical approaches, while N157 is identified only by the vertical approach. Window 5080 also produces 2 anomalies: N220 and N225, both of which are identified only by the horizontal approach. Windows 7700 and 14240 identify nodes N239 and N021 respectively. 

\begin{table}[!ht]
	\centering
	\caption{Amalgamation of horizontal and vertical anomalies for 4 windows.}
	\footnotesize
	\begin{tabular}{P{1cm}P{0.5cm}P{1.2cm}P{1.4cm}P{1.2cm}P{1.2cm}P{6cm}}
	\toprule
   Window & Node & Horizontal Anomaly & Vertical Anomaly  &  Horizontal  Anomaly Score   &  Vertical  Anomaly  Score  & Anomaly History (previous windows)\\
    \midrule
    2270 & N046 & $\qquad \checkmark$  & $\qquad \checkmark$ &   10 & 19.9 & 2269, 2268, 2267, \ldots\\
      & N157 & $\qquad$ --  & $\qquad \checkmark$ &   -- & 11.8 & 2269, 2268, 2267, \ldots\\
    \hdashline 
    5080 & N220 &  $\qquad \checkmark$ & $\qquad$ --   & 10 & -- &  5079, 5078, 5077, \ldots \\
      & N225 &  $\qquad \checkmark$ &  $\qquad$ --   & 10 & -- &  5079, 5078, 5077, \ldots \\
    \hdashline
     7700 & N239 & $\qquad$ --  & $\qquad \checkmark$  & -- & 10 & 7610, 7586, 7585, \ldots \\
      \hdashline
    14240 & N021 & $\qquad \checkmark$  & $\qquad \checkmark$ &  9.97  & 9.93 & 14239, 14238, 14237,  \ldots \\
	\bottomrule
	\end{tabular}
	\label{tab:amalgamation}
\end{table}

From Table~\ref{tab:amalgamation}, given Lookout's low false positive rates, it is clear that every anomalous node identified, even by a single approach, horizontal or vertical,  should be further investigated. We do this by comparing the results of Table~\ref{tab:amalgamation} with that of Table~\ref{tab:earlyanomalies}. Of the anomalies identified in window 2270, for example, N046 is a real anomalies -- we regard it as suspicious because it accesses the honeypot.  N157 is a false positive as it does not access the honeypot. Similarly, in window 5080, both N220 and N225 are real anomalies, as is the case with N239 and N021. 

\section{Insights and Discussion} \label{sec:insights}
Lookout is the preferred anomaly detection method employed in Honeyboost. As shown in Figure~\ref{fig:framework}, Honeyboost encompasses the complete framework, starting from the pre-processing and ending with amalgamating the anomalies that are found by vertical and horizontal approaches.   

\subsection{Tracking anomalies as they develop over time}

The amalgamation of the horizontal and vertical approaches lets us observe the  anomalies as they develop in different spaces. For example, a node may be identified as anomalous by the horizontal approach and later identified as anomalous by both approaches. The vertical approach identifies anomalies in each of the ARP, TCP and UDP spaces separately and then combines them. In contrast, the horizontal approach identifies anomalies in the combined feature space.

\begin{figure}[!ht]
     \centering
    \includegraphics{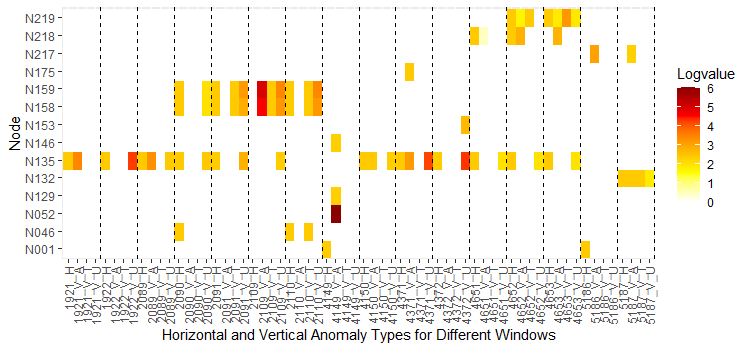}  
     \caption{Anomalies detected by the vertical and horizontal approaches in windows 1921, 1922, 2089, 2090, 2091, 2109, 2110, 4149, 4150, 4371, 4372, 4651, 4652, 4653, 5186 and 5187. As the vertical approach consists of each of the ARP, UDP and TCP spaces, anomalies identified separately in these spaces are  included. The vertical axis shows the anomalous nodes and the horizontal axis gives the windows and anomaly types. For example 1921\_H denotes horizontal anomalies identified in window 1921 and 1921\_V\_A denotes the vertical anomalies identified in the ARP space in window 1921. The colors indicate the logarithm of anomaly scores with darker shades denoting larger scores.    } 
     \label{fig:developinganomaly}
\end{figure} 

  Figure~\ref{fig:developinganomaly} shows the windows 1921, 1922, 2089, 2090, 2091, 2109, 2110, 4149, 4150, 4371, 4372, 4651, 4652, 4653, 5186 and 5187 and all  anomalies identified in these windows by Lookout. The $Y$ axis shows all nodes labeled as anomalous in these windows. The $X$ axis gives the different anomaly types sorted by windows. The anomalies identified in the horizontal and vertical approaches are differentiated with an \_H and \_V label, respectively. The vertical approach anomalies are further classified as ARP, TCP and UDP space anomalies. Thus, if an anomaly is identified by the vertical approach in the ARP space it is tagged with \_V\_A. Similarly, vertical TCP and UDP anomalies are tagged with \_V\_T and \_V\_U, respectively. The $X$ axis labels start with the window number followed by the anomaly type. Thus, 1921\_H denotes horizontal anomalies in window 1921. Similarly 1921\_V\_A, 1921\_V\_T and 1921\_V\_U  denotes vertical ARP, TCP and UDP anomalies in window 1921. The color of the cells denote the logarithm of the anomaly scores  with darker shades denoting larger scores. The windows are separated by dashed lines.  

We look at an example of a developing anomaly. In window 1921, N135 is the only node identified as anomalous, and is  identified as such by the horizontal  as well as the vertical approach, though only in the ARP space in the latter case.  It is then identified in the vertical UDP space, in window 1922, with a relatively high anomaly score as seen from the darker color.  It is further identified in window 2089, in the horizontal, as well as vertical ARP and UDP spaces. Again, in later time windows (4150, 4371),  it is identified as anomalous in the horizontal, and vertical ARP and UDP spaces. 

Nodes N158 and N159 are two further examples.  They are identified as anomalous in window 2090 by the horizontal approach, as well as in the vertical UDP space. These two nodes continue to increase in anomalousness in subsequent windows as seen by the darker shades of red. In window 2109, both nodes are identified by the vertical approach in the ARP, UDP and TCP spaces, with high anomaly scores. 

Other examples of developing anomalies in these set of windows are N218 and N219. Thus, the amalgamation of the horizontal and vertical approaches enables us to see the anomalies develop, assisting in the decision making process. Furthermore, Figure~\ref{fig:developinganomaly}  shows all anomalies identified by the two approaches in this set of windows. The maximum number of anomalies identified by Lookout in this set of windows is 4, occurring in windows 2090 and 4149. In comparison, the OCSVM identifies an average of 32.7 anomalies per window for this set of windows, with a minimum of 3 and a maximum of 58 nodes. Having a smaller number of anomalies increases clarity and helps  focus on the important aspects. This is another advantage of identifying anomalies sparingly.\\

\subsection{Tracking a node's anomalous nature}

\begin{figure}[!ht]
     \centering
    \includegraphics{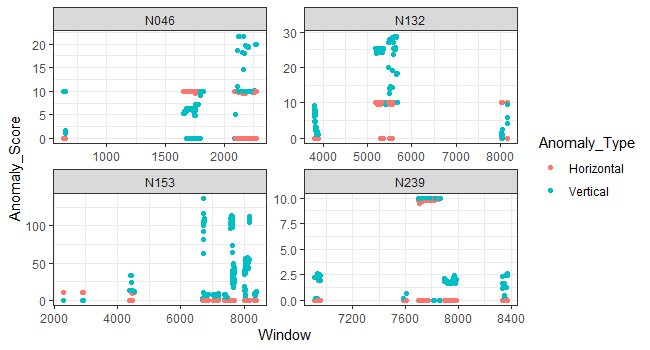}  
     \caption{ The horizontal and vertical anomaly scores for nodes N046, N132, N153 and N239 over all time windows. The $X$ axis of each graph displays the range of time windows for which these nodes are identified as anomalous starting from the window they were first found as anomalous, to the last window. For example, N132 does not get identified before window 3791. As such, windows before 3791 are not included in the graph for N132.   }
     \label{fig:fournodes}
\end{figure}

\begin{figure}[!ht]
     \centering
    \includegraphics{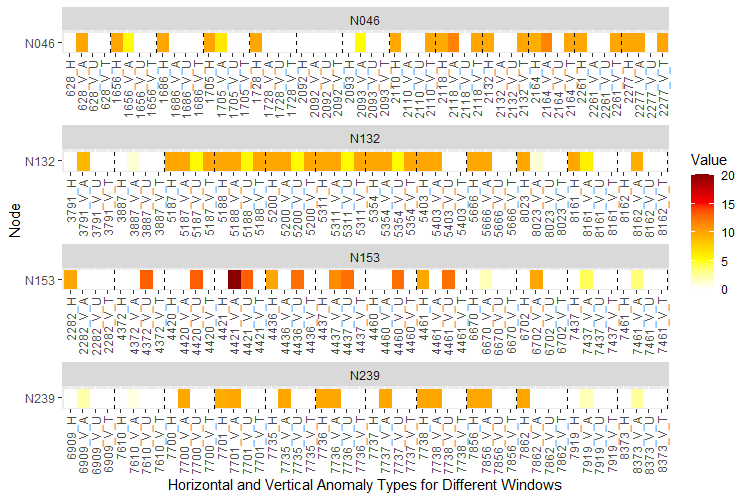}  
     \caption{Selected windows for nodes N046, N132, N153 and N239. The anomaly scores for the horizontal approach, and the vertical approach in the ARP, UDP and TCP spaces, are shown for each node with dashed lines showing window boundaries. We see the behavior of the nodes over time. } 
     \label{fig:developinganomalyindividual}
\end{figure}

Next, we focus on individual nodes instead of looking at all anomalies identified in a given window.  Figures~\ref{fig:fournodes} and~\ref{fig:developinganomalyindividual} show the behavior of nodes N046, N132, N153 and N239 over a range of windows along with their anomaly scores. From Figure~\ref{fig:developinganomalyindividual} we see that these nodes are identified as anomalous either by a single approach or by both approaches in sets of consecutive windows. For example, N046 is identified as anomalous in windows 628 to 650,  1656 to 1822 and in windows 2090 to 2277. We see a similar behavior by each of the other nodes. 

This observation is further validated in Figure~\ref{fig:developinganomalyindividual}, in which we explore the ARP, UDP and TCP anomaly spaces for the vertical approach in addition to the horizontal approach. We only show a subset of windows in Figure~\ref{fig:developinganomalyindividual} due to space constraints. Node N046 is first identified by the vertical approach in the ARP space in window 628. Later, in window 1656 it is identified as anomalous by the horizontal approach as well. 
This behavior continues for a while and in window 2118 it is identified by the horizontal approach and the vertical approach in the ARP and TCP spaces. It continues to be identified in multiple vertical approach spaces until window 2277. 

We see a similar pattern for node N132. Initially it is identified by the vertical approach in the ARP space in window 3791. In window 5187 it gets identified by the horizontal approach, and the vertical approach in the ARP, UDP and the TCP spaces. This continues for many windows and then it slowly dies down. Nodes N153 and N239 also display a similar pattern where they are identified by a single approach and then by both approaches in multiple spaces with higher anomaly scores. 

\subsection{Insights into anomalous behavior}\label{sec:insights1}
\begin{figure}[!ht]
     \centering
    \includegraphics[width=0.9\textwidth]{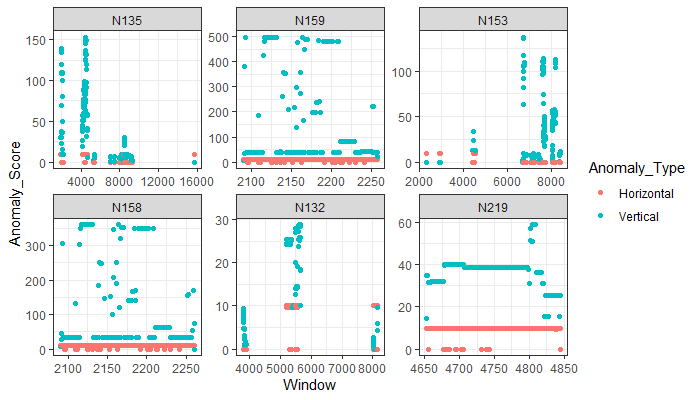}  
     \caption{The horizontal and vertical anomaly scores for the top 6 anomalous nodes accessing the honeypot, as per Table~\ref{tab:totalscores}.  } 
     \label{fig:top6anomalies}
\end{figure}

We now combine the results found thus far to gain insights into anomalous behavior. 
Table~\ref{tab:totalscores} lists the total anomaly score, the number of windows reporting the node as anomalous, and the average anomaly score per window, of the nodes accessing the honeypot. The total anomaly score for each node is the sum of the horizontal and vertical anomaly scores for the full time period. Figure~\ref{fig:top6anomalies} shows the vertical and horizontal anomaly scores for the top 6 anomalous nodes from this table.  We investigate these anomalies to gain better understanding. 

\begin{table}[!ht]
	\centering
	\caption{Nodes accessing the honeypot sorted by the total anomaly score. The number of windows reporting the node as anomalous as well as the average score per window are also given. }
	\footnotesize
	\begin{tabular}{p{1cm}p{2cm}P{2cm}P{2cm}P{2cm}P{2cm}}
	\toprule
   Node & Horizontal Anomaly Score Total &  Vertical Anomaly Score Total  &  Total Anomaly Score &  Number of Anomalous Windows & Average Score Per Window\\
   \midrule
    N135  &     6998  &  62473  &   69471   &  1351  &   51.4  \\
    N159  &     1487  &  30180  &   31666   & 169    &   187.0  \\
    N153  &      959  &  23902  &     24862 & 848    & 29.3 \\
    N158  &     1517  & 22235   &    23752  & 172    & 138.0  \\
    N132  &     4119  &  9305   &    13424  &  551   &  24.4 \\
    N219  &     1793  &  7072   &     8865  & 194    & 45.7\\
    N021  &     2615  &  5657   &     8272  & 766    & 10.8\\
    N155  &     1420  &  6002   &     7421  & 427    & 17.4\\
    N046  &     1965  &  3191   &     5157  & 360    & 14.3\\
    N039  &     2092  &  2297   &     4389  & 210    & 20.9\\
    N163  &     1060  &  2670   &      3730 & 152    & 24.5\\
    N239  &     1333  &  1899   &     3232  & 334    &   9.68\\
    N225  &     1070  &  1291   &     2361  & 247    &   9.56\\
    N220  &     1240  &     0   &      1240 & 124    & 10.0\\
    N171  &      940  &    18 &       958   & 98     & 9.78\\
	\bottomrule
	\end{tabular}
	\label{tab:totalscores}
\end{table}

\begin{enumerate}
    \item \textbf{N135}: This node suddenly makes a large number of ARP broadcasts (count 757, degree 257) and accesses the honeypot using UDP and TCP. The total length of the UDP packets per 5s time interval are unusually high for N135. The median value of the total length of the UDP packets, taking into account all nodes accessing the honeypot, is 28. For N135 the total UDP packet length ranges from 28 to 1047, with a median of 927. N135 has the highest total anomaly score because it is identified in 1351 windows. From Figure~\ref{fig:top6anomalies} we see that it displays a lot of activity for a long period of time. 
   
    \item \textbf{N159}: This node makes innocuous ARP broadcasts for the first 9 minutes of its communication. Then, it suddenly makes a large number of ARP broadcasts  (count 6760, degree 3384) and starts accessing the honeypot immediately after that, using 32 UDP ports in a single 5 second interval. After a  time lag of 1 hour, it accesses the honeypot using 999 TCP ports. From Figure~\ref{fig:top6anomalies} we see that the vertical anomaly scores in certain windows are much higher for N159 compared to other nodes.  Furthermore, it is only found anomalous in 169 windows. Thus, N159 has the highest average anomaly score per window. 
    
    \item \textbf{N153}: This node is found anomalous in 848 windows. It has a long span of activity as seen from Figure~\ref{fig:top6anomalies}. In terms of the ARP broadcasts, it has a higher count to degree ratio compared to most other nodes. 
    
    \item \textbf{N158}: This node has very large ARP counts for a long period of time, with a maximum ARP count of 6536 and degree 3262 for a given time interval. After these ARP broadcasts it accesses the honeypot using 32 UDP ports and 999 TCP ports.  
    \item \textbf{N132}: Again, we see extremely large ARP counts for multiple time intervals. The maximum ARP count for N132 is 22558 with degree 4416. It then accesses the honeypot using TCP and UDP protocols.  We find that N132 has set the PSH (push) and the URG (urgent) flags in some TCP packets that access the honeypot. The PSH flag informs the receiving node that the data should be pushed to the application layer immediately, and the URG flag informs the receiver that the data should be prioritized.  
    \item \textbf{N219}: Similar to N132, this node sets the PSH and URG flags for some TCP packets and accesses the honeypot using a large number of TCP ports. 
\end{enumerate}

Certain patterns emerge from this analysis. Some nodes such as N135 and N153 are identified as anomalous in a large number of windows. These nodes access the honeypot over a long period of time. In contrast, other nodes such as N219, access the honeypot over a small time interval and are identified as anomalous with high anomaly scores. 

Furthermore, we see different types of suspicious behavior:  unusually high number of ARP broadcasts by a node before accessing the honeypot followed by TCP or UDP packets targeted at the honeypot using a large number of ports; and nodes accessing the honeypot using  specific TCP flags that requests urgent attention and access to higher layers. 
Overall, it is clear that by combining the horizontal and vertical approaches, Honeyboost, via Lookout, prioritizes the nodes that need more inspection. 

\subsection{False Positives vs identifying anomalies}

Honeyboost identifies all nodes accessing the honeypot. In addition, it identifies most nodes before they access the honeypot. Of the identified anomalies, all do not access the honeypot, i.e. some are false positives. Table~\ref{tab:totalscoresall} shows the top 3 anomalous nodes, N001, N135, and N157,  identified by Honeyboost and their anomaly scores.  

\begin{table}[!ht]
	\centering
	\caption{Top 3 Honeyboost anomalies sorted by the total anomaly score. The number of anomalous windows and average score per window are also given.}
	\footnotesize
	\begin{tabular}{p{1cm}p{2cm}P{2cm}P{2cm}P{2cm}P{2cm}}
	\toprule
   Node & Horizontal Anomaly Score Total &  Vertical Anomaly Score Total  &  Total Anomaly Score &  Number of Anomalous Windows & Average Score Per Window\\
   \midrule
    N157  &      230  &  585540 &    585770 &   86   & 6811 \\ 
    N001  &    72178  &    5744 &     77921 &   7445 &  10.5 \\
    N135  &     6998  &  62473  &   69471   &  1351  &   51.4  \\
	\bottomrule
	\end{tabular}
	\label{tab:totalscoresall}
\end{table}

Of these 3 nodes, we discussed N135 in Section~\ref{sec:insights1}. The other 2 nodes, N157 and N001 do not access the honeypot. However, it is worth discussing their behavior. 

\begin{figure}[!ht]
     \centering
    \includegraphics[width=0.9\textwidth]{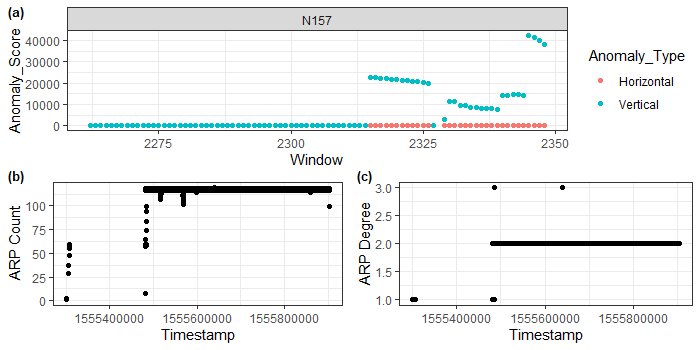}  %
     \caption{Behavior of Node N157: (a) Anomaly scores. (b) The ARP count over time. (c) The ARP degree  over time.}
     \label{fig:N157}
\end{figure}

Figure~\ref{fig:N157} shows the anomaly scores, ARP count and degree values for N157 over time. From Figure~\ref{fig:N157}(a), we see that the vertical anomaly scores for N157 are extremely large. Many anomaly scores are greater than 5000, and some scores are over 40,000. In fact, from Figure~\ref{fig:N157}(a) it is misleading to think that the horizontal score is zero for all windows. For some windows the horizontal score is 10, the maximum score for the horizontal approach, but this is dwarfed by the high values in the vertical axis. The vertical approach does not have a maximum score as a node can make multiple calls using ARP, UDP and TCP protocols and if many of them are identified anomalous the sum of the anomaly scores is considered for each window.  

The reason for high anomaly scores from the vertical approach is the large number of ARP calls with high count that are not balanced by high degree values as seen in Figures~\ref{fig:N157}(b) and (c). Starting from timestamp 1555482240, for every 60s interval N157 makes over 110 ARP broadcasts over 6381 minutes continuously as can be seen from the thick, black line in Figure~\ref{fig:N157}(b). It is rather unusual for a node to make such ARP broadcasts continuously for a long period of time. Especially, as it did not make such ARP broadcasts before that time. Therefore, this node needs  investigation. 

\begin{figure}[!ht]
     \centering
    \includegraphics[width=0.9\textwidth]{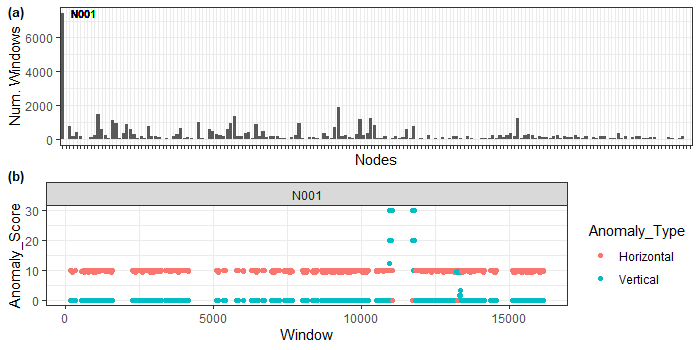}  %
     \caption{Behavior of Node N001: (a) Histogram of the nodes identified as anomalies in all time windows with N001 getting identified 7445 times. N001 is labeled in this histogram. (b) The horizontal and vertical anomaly scores over all windows for N001.}
     \label{fig:N001}
\end{figure}

Another node of interest is N001. Figure~\ref{fig:N001}(a) shows a histogram of the number of windows each node is found anomalous. N001 is found anomalous in 7445 windows and is the most frequently identified anomaly surpassing the other nodes by a large margin. Figure~\ref{fig:N001}(b) shows the anomaly scores for N001 for the horizontal and vertical approaches. On investigating the feature space, we find that this node has a relatively large total length of 73.4 in the ARP space (Section~\ref{sec:metam}, feature 6). It turns out this node has made 4635 ARP broadcasts at different time points in its first anomalous time window, all having count and degree values of 1, 2, or 3. Over the full time period, it makes 445,100 ARP broadcasts. Therefore, N001 is unusual because it makes a small number of ARP broadcasts extremely frequently over a long period of time. It might be a malfunctioning node needing investigation. We also note that node N001 would not have been identified by simply tagging the nodes making a large number of ARP broadcasts. 


\section{Conclusion}\label{sec:conclusions}
In this paper, we presented Honeyboost, a novel, hybrid framework consisting of two complementary approaches -- horizontal and vertical -- to enhance honeypot aided NAD. 
Both approaches use data fusion techniques to integrate attributes from diverse protocols. 
Our methodology does not suffer from a high false positive rate as we use an anomaly detection method called Lookout, which uses extreme value theory to identify anomalies. Furthermore, it operates totally unsupervised, alleviating the need for costly labeling procedures. Using our framework, we successfully identified anomalous nodes before they accessed the honeypot. 
In addition, we gained useful insights about the behavior of these nodes. Moreover, we identified some unusual behavior by nodes that do not access the honeypot. As future work, we plan to investigate network science and graph theory approaches to further boost early detection and classification of anomalies.
\bibliography{references}
\bibliographystyle{agsm}
%



\end{document}